# Imaging bulk and edge transport near the Dirac point in graphene moiré superlattices

Ziwei Dou[1], Sei Morikawa[2], Alessandro Cresti[3], Shu-Wei Wang[1], Charles G. Smith[1], Christos Melios[4], Olga Kazakova[4], Kenji Watanabe[5], Takashi Taniguchi[5], Satoru Masubuchi[2], Tomoki Machida[2,6], and Malcolm R. Connolly[1,†,*]

[1]Cavendish Laboratory, Department of Physics, University of Cambridge, CB3 0HE, UK
[2]Institute of Industrial Science, University of Tokyo, 4-6-1 Komaba, Meguro, Tokyo 153-8505, Japan
[3]Univ. Grenoble Alpes, CNRS, Grenoble INP, IMEP-LaHC, F-38000 Grenoble, France
[4]National Physical Laboratory, Hampton Rd, Teddington TW11 0LW, UK
[5]National Institute for Materials Science, 1-1 Namiki, Tsukuba 305-0044, Japan
[6]Institute for Nano Quantum Information Electronics, University of Tokyo, 4-6-1 Komaba, Meguro, Tokyo 153-8505, Japan
† Niels Bohr Institutet, University of Copenhagen, Universitetsparken 5, 2100 København Ø.
*Correspondence to: mrc61@cam.ac.uk.

**Abstract**

Van der Waals structures formed by aligning monolayer graphene with insulating layers of hexagonal boron nitride exhibit a moiré superlattice that is expected to break sublattice symmetry. Despite an energy gap of several tens of millielectron volts opening in the Dirac spectrum, electrical resistivity remains lower than expected at low temperature and varies between devices. While subgap states are likely to play a role in this behavior, their precise nature is unclear. We present a scanning gate microscopy study of moiré superlattice devices with comparable activation energy but with different charge disorder levels. In the device with higher charge impurity (~$10^{10}$ cm$^{-2}$) and lower resistivity (~10 kΩ) at the Dirac point we observe current flow along the graphene edges. Combined with simulations, our measurements suggest that enhanced edge doping is responsible for this effect. In addition, a device with low charge impurity (~$10^{9}$ cm$^{-2}$) and higher resistivity (~100 kΩ) shows subgap states in the bulk, consistent with the absence of shunting by edge currents.

**Introduction**

The mobility of charge carriers in two-dimensional (2D) crystals of graphene is enhanced by encapsulating them between atomically flat layers of hexagonal boron-nitride (hBN)[1]. Alongside improvements in device quality has come a rich array of physics arising from the van der Waals interaction between the 2D layers. Moiré superlattices[2, 3], for instance, are formed by aligning the graphene lattice within a few degrees of the hBN [Fig. 1(a)]. Strain-relaxation[4, 5] combined with different crystal potentials from the boron and nitrogen atoms in the hBN breaks the symmetry between the graphene sublattices and opens a gap in the quasiparticle density of states at the Dirac point[2, 3, 4, 6]. An energy gap of the order of 10 meV is typically extracted from thermally activated transport near the charge neutrality point (CNP)[3, 4, 6] and shows the expected functional dependence on the moiré wavelength $\lambda_M$[3, 7]. Activated transport is observed in the high temperature regime (> ~50K), but at lower temperature the resistivity at the CNP shows a much weaker temperature dependence[3, 4, 6]. Similar behavior was observed previously in gapped bilayer graphene (BLG)[8, 9] and attributed to the transition from thermal excitation over a bandgap to hopping conduction in the bulk via low energy states induced by charge disorder[9]. Transport via edge states could also shunt the insulating bulk and lead to resistance saturation[10-12], but no clear signatures of edge transport near the CNP were observed in conventional low-mobility graphene on $SiO_2$ substrates[13]. Edge states are, however, more likely to play a role in high-mobility graphene where bulk charge disorder is reduced. The distinction between edge and bulk transport in gapped graphene may also influence topological valley Hall effects (VHE), both in gapped moiré MLG[6, 14] and in the dual-gated BLG devices[14-16], where non-zero Berry curvature is present due to broken sublattice symmetry[17]. Understanding the current distribution in gapped graphene devices near the CNP is therefore a high concern. Indeed, recent Josephson interferometry in very short (~100 nm) graphene junctions between superconducting Nb contacts shows a transition from bulk-to-edge transport occurs as the energy approaches the CNP in both gapped BLG and MLG[12, 18].

In this work, we use scanning gate microscopy (SGM) to examine the current distribution in micron-sized MLG moiré superlattices. We present data taken from two devices with comparable activation energies ($E_g/2$ ~14 meV) but different levels of disorder-induced energy broadening, $E_d$. In device D1, $E_d > E_g/2$, and the resistivity at the CNP saturates at ~10 kΩ, comparable to the $h/e^2$, whereas in device D2, $E_d < E_g/2$, and the CNP resistivity (~ 100 kΩ). Our SGM images show a clear response at the edges in D1 near the CNP. Despite its lower disorder, the SGM response in D2 is exclusively in the bulk. Using tight-binding simulations as a guide, we are able to explain our results by enhanced doping at the edges of D1. Our results therefore suggest that, while edge transport plays an important role in shunting the gapped bulk in MLG moiré superlattice devices, it probably originates from external factors such as disorder and inhomogeneous doping rather than universal properties of gapped Dirac spectrum[10, 12].

**Experiment**

Our MLG encapsulated moiré superlattice devices [Fig. 1(a)] are fabricated by mechanical exfoliation and the "pick-up" transfer technique of atomic layers[19]. The MLG is sandwiched between a 50 nm top layer hBN and a 70 nm bottom layer hBN, and the whole structure is placed on a ~290 nm $SiO_2$ formed on a doped Si substrate used for applying a back-gate voltage ($V_{BG}$). The devices are then patterned into Hall bars and Cr/Pd/Au ohmic contacts are made via one-dimensional edge contacts[19]. An optical image of the device and circuit schematic is shown in Fig. 1(a). We use standard low-frequency lock-in techniques to determine the four-terminal resistance $R_{4T}$ = $V_{xx}/I_{SD}$. For D1, we first confirm the presence of a moiré superlattice by observing secondary Dirac points (SDPs) in $R_{4T}(n)$, a well-known signature resulting from the modified bandstructure of moiré MLG[3] [Fig. 1(b)]. The carrier density $n$ is calculated by $C_{BG}(V_{BG}-V_{CNP})$ where $C_{BG}$ = 1.124 × 10$^{-4}$ F/m$^2$ is the capacitance per area to the backgate extracted from quantum Hall measurements, and $V_{CNP}$ = 1.25V is the back-gate voltage at the charge neutrality point. From the SDP carrier densities we estimate $\lambda_M$ of ~ 11 nm [2,3,7]. Fig. 1(c) shows the resistivity $\rho$ (calculated from $R_{4T}$ ~ $\rho L/W$, where L = 11.5 μm and W = 1.3 μm are the length and width, respectively) measured at $T$ = 4 K as a function of $n$. From the full-width-half-maximum (FWHM) of the Dirac peak (~4×10$^{10}$ cm$^{-2}$), we estimate disorder-induced broadening of ~24 meV[3,4,6], comparable to other MLG-hBN studies[3,4,6,12]. Fig. 1(d) shows the temperature dependence of $\rho$ in the range 4 K - 200 K. The maximum resistivity at the CNP ($\rho^{CNP}$) is extracted from back-gate sweeps performed at each temperature and plotted against inverse temperature, resulting in an Arrhenius plot. We identify three regimes: (I) at high temperature (>70 K) the data are well described by a simple model for thermally activated transport across an energy gap $E_g$, $\rho^{CNP} \sim exp(\frac{E_g}{2kT})$, where $k$ is the Boltzmann constant. By fitting the data to this expression (red dashed line), we extract $E_g/2$ ~ 160 K (~ 14 meV), consistent with previous work[3,4,6] and the gap expected from the estimated $\lambda_M$[3]. (II) At intermediate temperatures (grey hatched lines) the weak temperature dependence is consistent with variable range hopping, and by fitting the resistivity to $\rho \sim exp[(T_0/T)^{1/3}]$, we deduce $T_0$~ 2 K[9,20]. (III) At low temperature, the resistivity saturates at ~ 10 kΩ, comparable to the resistance quantum $h/e^2$ and consistent with earlier works[3,4].

In an effort to resolve the microscopic mechanism behind the flattening of $\rho^{CNP}(T)$ in regime (III), we use SGM at $T$ ~ 4 K. SGM is a well-established technique for visualizing local variations in the electronic properties of 2DESs[21-27] and involves monitoring the conductance of a mesoscopic device while scanning a sharp metallic tip over its surface. Fig. 2(a) illustrates the SGM setup used in our experiments, where a constant current ($I_{SD}$ = 100 nA) is driven between the two ends of the superlattice while measuring the voltage $V_{xx}$. The tip is lifted ~100 nm above the top hBN and dc-biased with voltage $V_T$. The potential difference between the tip and the graphene modifies the carrier density under the tip, locally shifting the carrier density and Fermi level relative to the Dirac point. Figs. 2(b) and 2(c) shows two typical wide-area SGM images of the four-terminal resistance $R_{4T}$ taken at high hole density ($n \sim -2.5\times10^{10}$ cm$^{-2}$) and near the CNP ($n \sim 0$), respectively, with device edges marked in the black dotted lines. The scan area is also drawn in Fig. 2(a) as the red rectangle. In the hole-doped regime [Fig. 2(b), upper image], the resistance is reduced by ~ 500 Ω when the tip is located within the region highlighted by the dashed oval circle. This is consistent with a long-range gating effect that increases the hole density in the channel ($V_T$ = -0.5V) and reflects the cylindrical symmetry of the tip potential. When the tip is directly over the channel, however, $R_{4T}$ increases on average by ~ 1 kΩ relative to the background. We attribute this to a spatially dependent contact potential $V_0$, a well-known effect in SGM[21]. When the tip is away from the channel, $V_0$ is ~ 0 V and the effective tip potential $V_T^* = V_T-V_0 = $ -0.5V, reducing the resistance of the hole doped channel and producing the feature within the oval dashed line. When the tip is over the channel, $V_0$ is measured to be ~ -1 V, and $V_T^*$ = 0.5V, giving rise to the observed resistance enhancement (see Supplementary Information Sec. VI). With the global doping level at the CNP [Fig. 2(c), bottom image], a ~ 1 kΩ enhancement in $R_{4T}$ is seen only when the tip is over the edges of the flake [see plot of averaged $R_{4T}$ (y) in Fig. 2(c)], which cannot be explained by a spatially dependent contact potential (see Supplementary Information, Sec. VI).

We investigate more closely by capturing a sequence of high-resolution scans at different carrier densities over the range from -2.1 ×10$^{10}$ cm$^{-2}$ to 0.7 ×10$^{10}$ cm$^{-2}$ (note the unequal carrier density intervals denoted in the caption), and plot the SGM micrographs in Figs. 2(d). The scanned region is outlined in Fig. 2(b) by the black rectangular box and the device edges are marked as black dotted lines. The "hotspots" (defined here as an isolated region of tip position where the resistance is higher than the background) that are initially over the bulk channel split laterally and migrate towards the edges. At the CNP some hotspots fragment further along the edge (see the dashed circles). To track the bulk to edge transition we perform scanning gate spectroscopy (SGS), which involves stepping the tip position along a line and sweeping the global carrier density at each point. Plotting the resulting $R_{4T}(n,y)$ allows us to determine the evolution of a particular feature with both high energy and high spatial resolution in one dimension. Since we are interested in the bulk to edge transition we choose the line [drawn in Fig. 2(d)] that crosses over both edges. The result is plotted in Figs. 2(e) in five panels. Note that each panel is plotted over a different range in order not to mask the ~1 kΩ modulation by the larger ~ 50 kΩ increase in the background resistance with decreasing $n$. As expected from the previous micrographs, away from the CNP the tip increases the resistance only in the bulk. Such bulk response continuously opens until it reaches the device edge at the CNP, similar to the hotspot features. The data are asymmetric about the Dirac point, and the bulk-edge transition predominantly occurs for hole doping. While in Fig. 2(e) a feature resembling the transition back to bulk response is seen as the device becomes electron doped, such features disappear altogether with more positive tip bias $V_T^*$ ~ +1.5 V in Figs. 2(f), where the bulk-edge transition occurs for hole doping only.

Before discussing the edge response in more detail, we examine a second device (D2) with similar properties but with ~3 times lower bulk disorder level. Fig. 3(a) shows an optical image of the device along with the two-terminal measurement schematic for SGM. We first check the transport properties of D2 using the same four-terminal measurement as in Fig.1 on D1. Figs. 3(b)-(d) show the Dirac peak, Arrhenius plot of the resistivity at CNP, and the SDPs, respectively. From Fig. 3(b), the charge carrier density disorder of D2 is ~ $0.4 \times 10^{10}$ cm$^{-2}$ (or ~ 7 meV), a factor of three lower than D1 (plotted in reference as the dashed line) and is comparable with the best quality reported in literature[3, 4, 6]. The temperature dependence [Fig. 3(c)] shows an activated region with similar transport gap ~ 28 meV and plateau at low temperature. Fig. 3(d) confirms the moiré superlattice of the sample with SDPs translated to similar moiré wavelength ~ 11 nm. Therefore, D2 has the same gap size as D1 but with lower bulk disorder. From earlier proposals[10, 12, 18], edge transport is expected to be more pronounced in D2. To test this we perform the equivalent SGM and SGS using two-terminal measurements [Fig. 3(a)]. Figs. 3(e) show a series of SGM images of $R_{2T}$ = $V_{SD}/I_{SD}$ with constant voltage ($V_{SD}$ = 1 mV) taken with the same scan conditions as Fig. 2 at different values of $n$ around the CNP. We do not observe an edge response, but only a few regions in the bulk where the tip can affect transport [see the dashed circles in Fig. 3(e)]. We follow the same method employed for D1 and perform SGS along the *y* direction perpendicular to the edge. The spectroscopy plotted in Fig. 3(f) shows a weak gating effect that produces an overall shift in the back-gate voltage of the CNP, but with no increase in $R_{2T}$ at the edge near the CNP [$R_{2T}(n)$ without the tip is plotted in Fig. 3(g) for reference]. The strong ~ MΩ variations due to the tip (as large as ~ 50 % of the total $R_{2T}$) are clearly concentrated in the bulk.

We perform SGS by moving the tip along a line, this time parallel to the channel [arrow along *x* in Fig. 3(e)] and intersecting the sites marked by the dashed circles. The spectroscopy is plotted in Fig. 3(h). When the tip is in the vicinity of each site, a peak in $R_{2T}(n)$ appears at higher *n*. This peak splits from the main Dirac point and follows a Lorentzian-shaped trajectory as a function of *x*. To emphasize these trajectories we take the derivative $dR_{2T}/dn$ in the area marked by the white box in Fig. 3(h), and plot the result in Fig. 3(i). The trajectories can be fitted by Lorentzians with FWHM of ~0.5 μm and origins separated by *Δx* ~ 1 μm. Line sections [Fig. 3(i)] show that height of each peak is comparable to the amount by which the main Dirac peak ($P_1$) is reduced (~ 1 MΩ), suggesting the sites add in series to yield the total resistance at the DP. We confirmed that sites are not associated with inhomogeneous strain in the sample[28] by checking the 2D peak width using scanning Raman spectroscopy (see Supplementary Information Sec. IX). Despite its lower bulk disorder, therefore, D2 does not show pronounced edge conduction[10, 12] and the transport appears mediated via subgap states in the bulk. The absence of edge shunting is also noteworthy, given that the resistivity at CNP plateaus higher than $h/e^2$ [12]. One natural explanation is that transport is via a series of electron-hole puddles forming a bulk percolation path[13]. High quality devices typically have ~ 100 nm size puddles[29, 30], consistent with a small number of conducting sites separated by insulating regions.

**Discussion**

Several mechanisms could give rise to the observed edge transport in D1. Possible scenarios include zigzag edge states without[11] or with energy dispersion under more accurate Hamiltonians[31], and states with enhanced localization length due to nontrivial bulk band topology from the broken sublattice symmetry[10, 12]. However, since these states are not topologically protected from inter-valley scattering, they localize easily and do not participate in transport[10, 32, 33]. The absence of edge response in D2 despite the lower disorder also suggests the resistivity at the DP is governed by extrinsic factors[10]. Two candidates are field-focusing[34, 35] and chemical doping[36]. We rule out the former as it is stronger away from the CNP due to higher screening ability at high carrier density[34, 35]. Moreover, in D1 the CNP is at $V_{BG}$ ~ +1 V. Field focusing should thus induce additional *n*-type carriers near the edge[34] and a positive applied $V_T^*$ should decrease the resistance, which is contrary to what we observe experimentally. The most likely mechanism for our observations of D1 is thus chemical doping at the edges. To test this scenario we perform tight-binding simulations on a 2 μm × 1 μm graphene lattice with a staggered sublattice potential ±Δ = ± 7.5 meV to break sublattice symmetry and to simulate the moiré-induced gap[17] [Fig. 4(a)]. Short-range random edge roughness is imposed on perfect zig-zag edges with (4,1) chirality to simulate localization of the edge states in realistic devices, and a long-ranged bulk disorder is included through randomized Gaussian potential fluctuations with maximum strength of 10 meV[37], whose strength relative to the gap size is in keeping with D1. We use a parabolic profile with 300 nm width and +30 meV amplitude to simulate the enhanced edge potential estimated from the spectroscopic measurements (see Supplementary Information Sec.V). Figs. 4(b) and (c) show the local spectral current map $j_0(x, y)$ of the system away from and at the charge neutrality point, respectively. At $E_F$ = -12.5 meV [Fig. 4(b)] the bulk is *p*-type, leading to bulk conduction. At $E_F$ = 0 [Fig. 4(c)], bulk transport is gapped out and the current flows at the edges. Note the edge current is present despite the edge roughness. To simulate the SGS experiment, we model the tip by a Lorentzian potential with 200 nm FWHM. We choose a maximum tip-induced potential energy $U_T$ = - 20 meV, estimated from the effective tip bias $V_T^*$=+1.5V and a parallel plate capacitor model based on a tip-sample distance ~ 100 nm. Fig. 4(d) show the simulated SGS with the channel resistance $R_{2T}$ calculated from the transmission coefficient. The SGS [Fig. 4(d)] is plotted in five panels with different color scales similar to Fig. 2. The tip passes over the two edges of the channel at $y_T$ = 0 and 1 μm ($x_T$ = 1 μm) and the energy gap (± 7.5 meV) is marked by the dashed vertical lines. This plot captures some key features of the experimental SGS in Fig. 2, namely a clear enhancement of the resistance appears at the edges at around -12.5 meV in the hole-doped regime, and diminishes as $E_F$ is tuned away from CNP.

To illustrate how the above SGM features relate to the enhanced doping at the edges, in Fig. 4(e) and (f) we plot how the underlying current distributions $j_0(x,y)$ in Fig. 4(b) and (c) change in the presence of the tip, $\Delta j = j(x,y) - j_0(x,y)$. At the edge of the flake, the bandgap (the region between the red and blue solid lines) is lifted to higher energy [Fig. 4(e), right]. Hence, when $E_F$ = -12.5 meV [blue circle in Fig. 4(d)], the edge has a higher hole density than the bulk. Although the current is enhanced along the edges, the effect of the negative tip potential energy is weak because the bulk is conductive. The simulated $\Delta j$ map [Fig. 4(e)] consequently displays weak ambipolar modulations of the current and there is no enhancement at the edges. By contrast, when $E_F$ = 0, the $\Delta j$ map [Fig. 4(f)] shows a strong response along the edge because the conduction and valence bands are lowered and the hole density reduces [see bandstructure, Fig. 4(f)]. This gaps out the current flow at the edges and increases the resistance. In the inset of Fig. 4(f), the same $\Delta j$ map is plotted with positive tip potential. In this case, the edge current is locally enhanced and only weakly affects the global resistance, consistent with the absence of an SGS response. The fact that we observe an edge response only with positively biased tip (or negative tip potential energy) is further consistent with enhanced *p*-type doping (see Supplementary Information Sec. VII). Within this framework potential disorder along the channel creates the hotspots in Fig. 2, while their migration towards the edge reflects the transverse potential profile caused by edge doping. Note, however, that the measured edge response is weaker than in the simulations. It is likely that the bulk is more conducting in the real device due to disorder, which is nevertheless not strong enough to totally mask edge transport[13]. The precise microscopic origin of the enhanced edge potential is also unclear, but may arise from the trapped molecules (e.g. water) between the bottom hBN and the $SiO_2$ substrate, consistent with the observed hole doping in SGS[36].

In summary, we have used SGM to study subgap transport near the CNP of gapped moiré MLG devices. In one device we observe a clear transition from bulk to edge response as the Fermi level approaches the CNP, suggesting currents flow at the edges. Guided by numerical simulations, we showed that transport can be explained by hole doping near the edges. The absence of such edge conduction in a much cleaner device also suggests that edge states[10, 12] do not always shunt the bulk. These observations may serve as a starting point for improving the insulating behavior using local gates to compensate doping at the edges. Direct SGM imaging of bulk and edge currents implicated in the VHE [38, 39] and at the SDPs would be a natural extension of this work. Moreover, our results show how SGM could be used to examine transport in a wide range of low-dimensional materials, such as topological insulators[40] and superconductor-semiconductor hybrids[41], where differentiating between trivial and topological edge modes could prove important for proposed applications in quantum computing[42] and low power electronics.

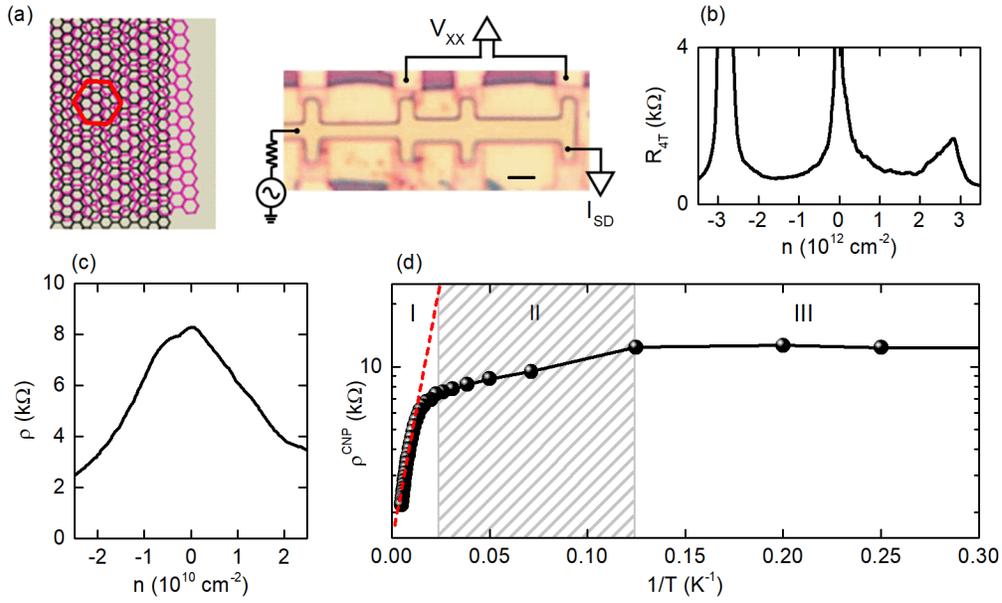

**Figure 1.** Transport characterization of MLG moiré Hall bar device (D1).

(a) Left: moiré superlattice formed by aligning monolayer graphene (black) with h-BN substrate (magenta). The red hexagon depicts a moiré unit cell. Right: Optical image of the device with a schematic of the measurement setup (scale bar 1 μm). (b) Four-terminal longitudinal resistance $R_{4T} = V_{xx}/I_{SD}$ as a function of $n$ showing secondary Dirac points ($T$ = 4 K). (c) Resistivity as a function of carrier density ($T$ = 4 K). (d) Arrhenius plot of $\rho$ at the CNP as a function of $1/T$. Three regions are seen: (I) exponential dependence of $\rho$ on $1/T$ at high temperature; (II) variable-range hopping at immediate temperature (hatched grey); (III) constant $\rho$ at low temperature.

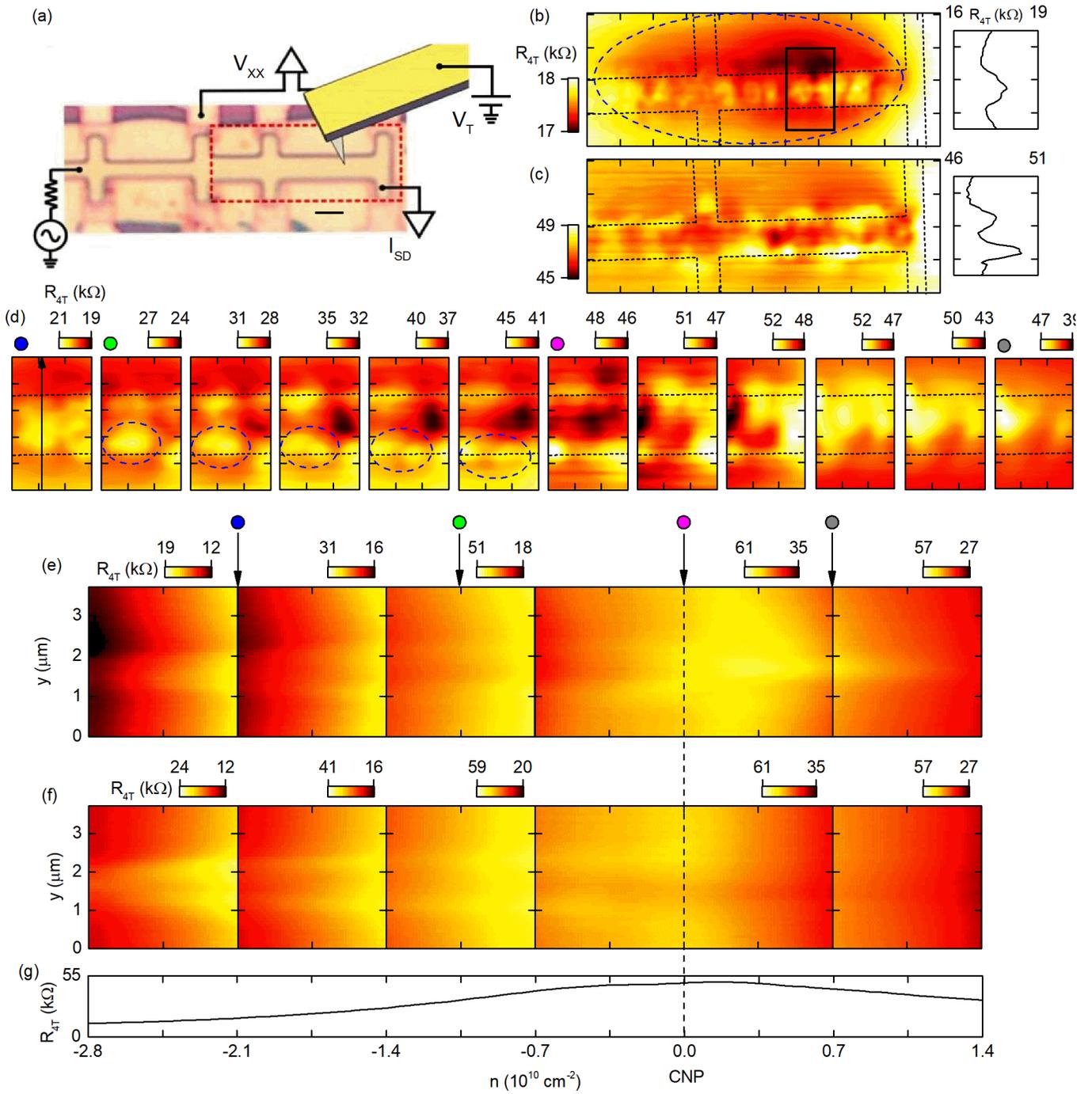

**Figure 2.** Scanning gate microscopy of device D1 (*T* = 4 K).

(a) Optical image of the device and measurement setup used for SGM. A dc biased AFM tip ($V_T^*$ ~ +0.5V, see main text) is lifted ~ 100 nm above the top hBN layer and locally gates the graphene. The scanned area in (b) and (c) are marked by the red dashed rectangle (tip not to scale, scale bar: 1 μm ). SGM image of $R_{4T} = V_{XX}/I_{SD}$ as a function of tip position when globally (b) $n \sim -2.5\times10^{10}$ cm$^{-2}$ and (c) $n \sim 0$ cm$^{-2}$. Edges of the device are outlined by dotted lines and each grid represents 1 μm. Blue dashed oval depicts the region to the long-range tip gating effect. Insets show $R_{4T}(y)$ averaged over all $x$. (d) Sequence of higher-resolution SGM images of $R_{4T}$ in the area marked by the black rectangle in (b). The images are taken at different $n$: -2.11×10$^{10}$ cm$^{-2}$ (blue); -1.05×10$^{10}$ cm$^{-2}$ to ~0 with ~ 0.21×10$^{10}$ cm$^{-2}$ intervals (green to magenta); 0.14×10$^{10}$ to 0.70×10$^{10}$ cm$^{-2}$ with ~ 0.14×10$^{10}$ cm$^{-2}$ intervals (magenta to grey). The device edges are outlined with dotted lines; each grid represents 0.5 μm. Evolution of a "hotspots" is marked by blue dashed circle. (e) SGS as a function of $n$ along the black arrow in (d). The SGS is plotted in five panels with different color scales for clarity. Corresponding values of $n$ for the images in (d) are marked with colored circles. (f) Same SGS as (e) but with $V_T^* \sim +1.5$V. (g) Reference $R_{4T}(n)$ without the tip. Plots in (e), (f), and (g) share the same horizontal axis. The CNP is marked by the vertical dashed line.

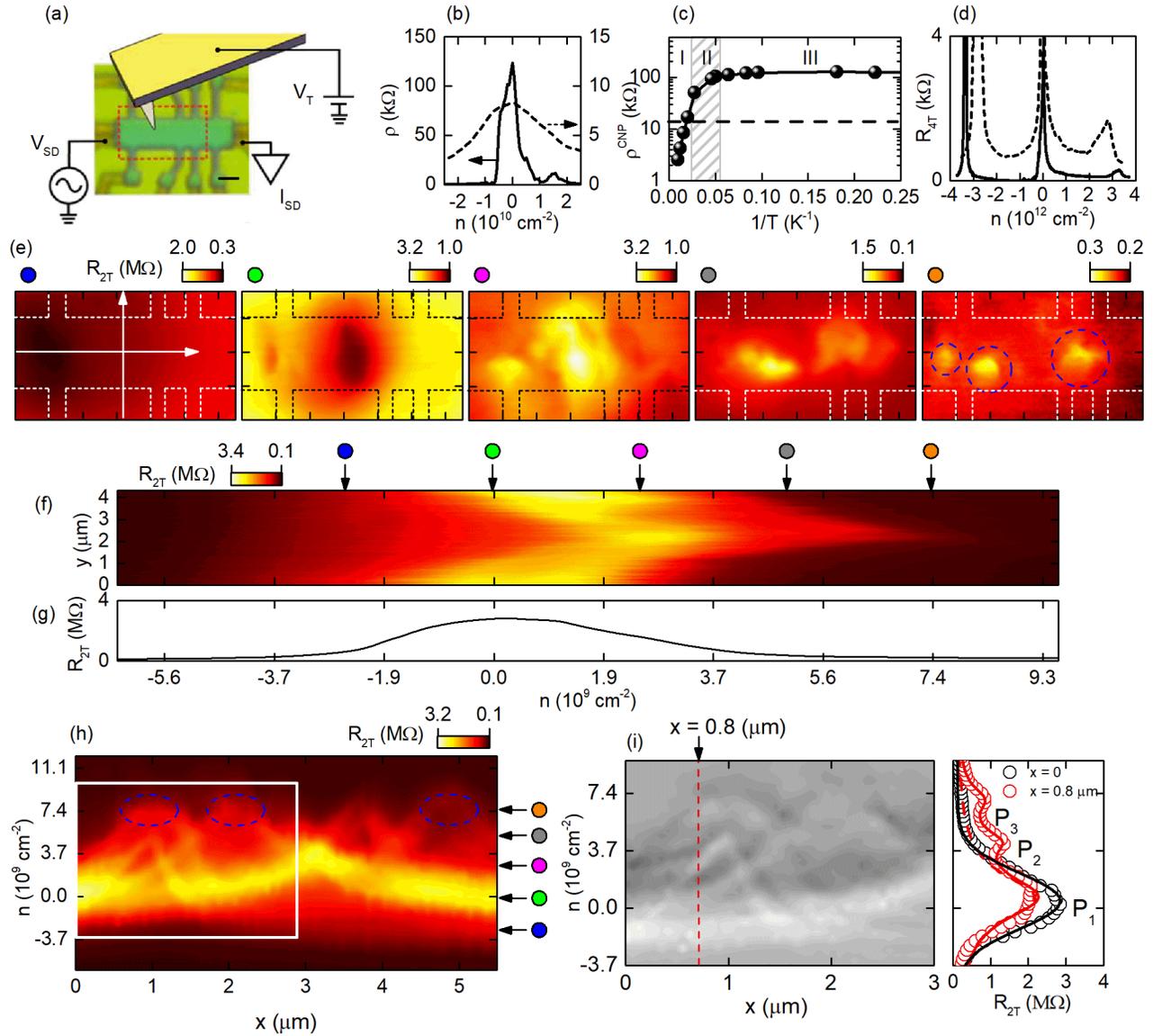

**Figure 3.** Scanning gate microscopy of device D2 ($T$ = 4 K).

(a) Optical image of the device and measurement setup used for SGM (tip not to scale, scale bar: 1 µm). Scanned area is marked by the red dashed rectangle. (b) Comparison between the resistivity $\rho(n)$ ($T$ = 4 K) of D2 (solid) and D1 (dashed), showing a narrower FWHM and ~10 times higher resistivity at the charge neutrality point for D2. (c) Arrhenius plot of $\rho$ at the CNP as a function of $1/T$. Three regions similar to D1 (Fig. 1) can be identified. Dashed line marks the low-temperature resistivity of D1. (d) $R_{4T}(n)$ plotted over a wider range of $n$ to show the SDPs for D2 (solid) and D1 (dashed) ($T$ =4 K). (e) Sequence of SGM images of $R_{2T} = V_{SD}/I_{SD}$ captured at equal $n$ intervals between -2.5×10$^9$ and 7.4×10$^9$ cm$^{-2}$. Localized sites of higher resistance are marked by the blue dashed [see (h) for corresponding spectroscopy]. The device boundaries are marked with dotted lines; each grid represents 1 μm. (f) SGS as a function of $n$ and distance $y$ along the vertical white arrow line drawn in (e). Corresponding values of $n$ for the images in (e) are marked with colored circles. (g) Reference $R_{2T}(n)$ without the tip. (h) SGS as a function of $n$ and distance $x$ along the horizontal white arrow in (e). The peaks of the Lorentzian trajectories associated with the sites in (e) are marked by blue dashed circle. (i) Differentiation of (h) in the $n$ direction in the region shown by white box in (h). Side: Line sections of $R_{2T}(n)$ along the red dashed line ($x$ =0.8 μm) (red circles) and along $x$ = 0 (black circles), with the Lorentzian fits plotted as the solid lines

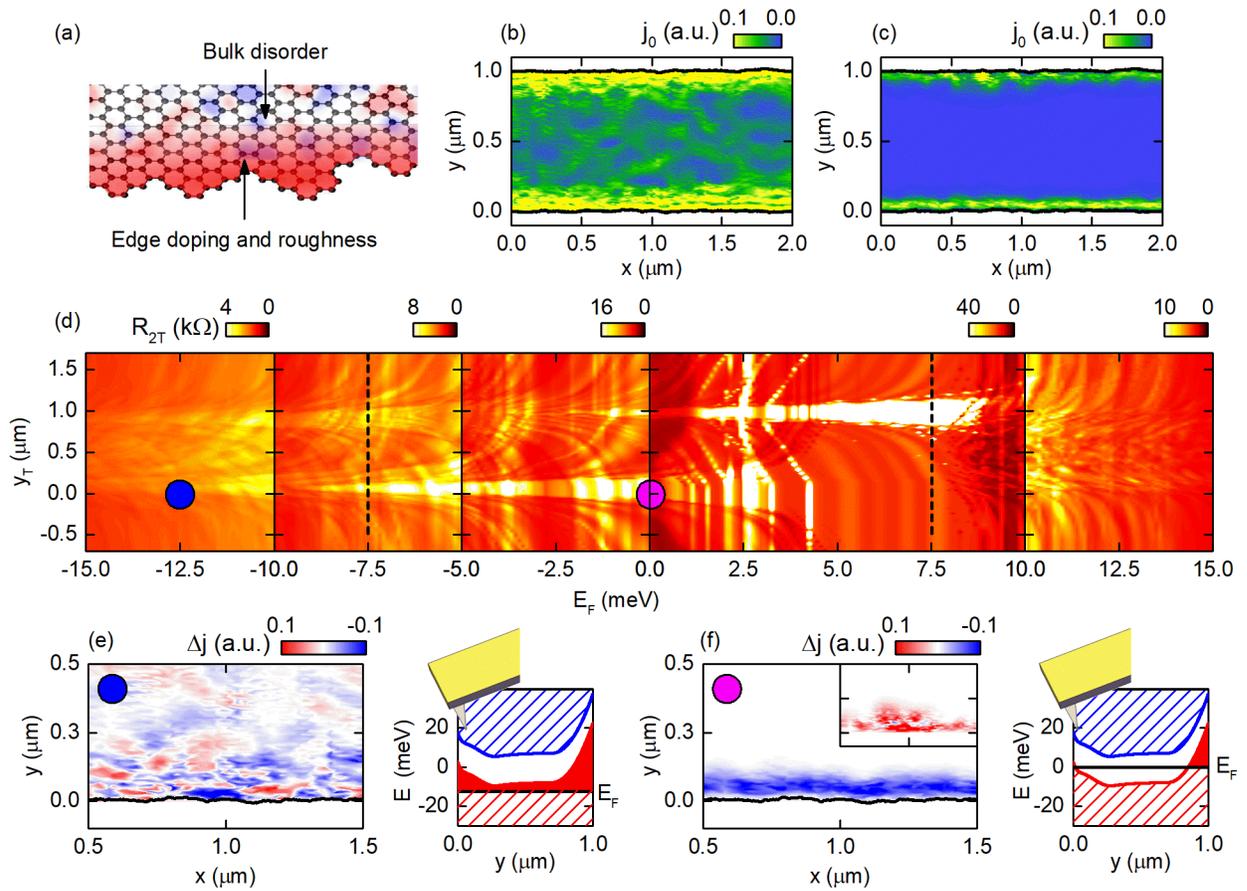

**Figure 4.** Simulated scanning gate spectroscopy.

(a) Cartoon showing the simulation setup. The gapped graphene system size is 2 μm × 1 μm, with edge potential, random edge roughness, and long-range potential disorder in the bulk. The SGM tip is simulated by a Lorentzian potential. Simulated local spectral current density map without the tip at (b) $E_F$ = -12.5 meV and (c) $E_F$ = 0 meV. Edges are highlighted by black lines. (d) Simulated SGS as a function of tip position and $E_F$ along a line crossing the edges at $y_T$ = 0 and 1 μm. The SGS is plotted in five panels in different color scales. The sublattice-induced gap in the Dirac spectrum at ±7.5 meV is marked by the black dashed lines. The tip-induced potential energy is $U_T$ = -20 meV. (e) Change in spectral current density with the tip potential centered at the edge $y_T$ = 0 and $E_F$ = -12.5 meV. The map corresponds to the blue filled circle in (d). Right: bandstructure as a function of $y$ across the graphene when the tip is at $y$=0. The empty conduction and valence bands are marked by blue and red hatched lines, respectively. The global Fermi level is marked by the horizontal black line and the filled valence band is filled in red. The gap is pushed towards higher $E$ by edge potential and is lowered near the edge $y = 0$ by the tip. (f) Same as (e) but with $E_F$ = 0 meV corresponding to the magenta circle in (d). Inset: same map but with $U_T$ = +20 meV. Right: bandstructure as a function of $y$ across the graphene when the tip is at $y=0$ and $E_F$ = 0 meV.

**Author Contributions**

*Z. D. conducted the low temperature transport measurement and all the low temperature scanning gate and spectroscopy measurement; S. Morikawa and S. Masubuchi prepared the stamped van der Waals heterostructure and fabricated the devices with supervision from T. M. and prepared them for the microscope with input from S. W. W.; Z. D. and M. R. C. conceived the idea, designed the experiments, analysed the data and wrote the manuscript with useful discussions from C. G. S., A. C., S. Morikawa, S. Masubuchi and T. M.; A. C. performed the tight-binding simulations; Ch. M and O. K. performed and supervised the Raman spectroscopy. K. W. and T. T. provided the h-BN material; M. R. C. and T. M. supervised the research. All authors have given approval to the final version of the manuscript.*

**Acknowledgements**

This work was partly supported by EPSRC EP/L020963/1, JST CREST Grant Numbers JPMJCR15F3 and JSPS KAKENHI Grant Numbers JP25107003, JP25107004, JP26248061, JP15H01010, JP16H00982. Additional data related to this publication is available at the University of Cambridge data repository.